# Microplastic Detection in Soil and Water Using Resonance Microwave Spectroscopy

Oleksandr Malyuskin

***Abstract*—** A feasibility study of microplastic detection and quantification in soil and water using resonance microwave reflectometry is carried out using artificially created samples with high volumetric concentration of microplastic with 50μm-0.5mm particles size. A mathematical model expressing microplastic concentration in soil and water as a linear function of the measured $S_{11}$ resonance frequency shift and relative permittivity contrast is developed and is found to be in a very good agreement with the experimental data. Next, this model is applied to find the best achievable theoretical resolution of microplastic concentration in the natural environment using microwave sensing technology which is shown to be at around 100ppm (parts-per-million) level in the linear signal detection regime. It is demonstrated that the best achievable level of microplastic contaminant resolution depends on the sensor probe Q-factor and sensitivity of the microwave receiver. The bound for the achievable contaminant concentration resolution is found in the analytical form for high-Q resonance microwave sensors of arbitrary geometry. Even though several well-established protocols based on optical, infrared and X-ray spectroscopy are currently being used for microplastic detection in the natural environment, microwave spectroscopy could offer additional possibilities, especially for low-cost, real-time in-situ microplastic detection in diverse environmental conditions outside of the laboratory space.

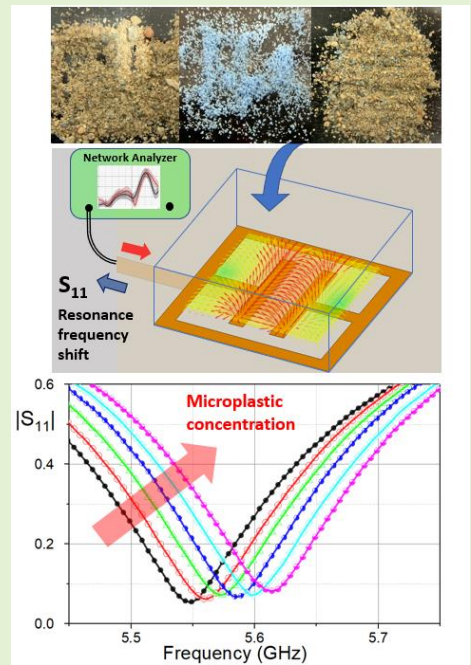

## I. Introduction

MICROPLASTICS are diverse polymer-based particles of characteristic size below 5mm that were found in abundance in natural environment including marine, fresh and drinking water, wastewater, soils, food and air [1]-[5]. Microplastic pollution is a growing environmental problem due to the current scale of global plastic production, estimated as 320 million tons per year [6] insufficient (9%-14%) recycling of plastic waste and relatively short life-span of plastic products, half of which are estimated to turn into trash in less than one year [7]. Eventually, more than 80% of manufactured plastic that has not been recycled ends up in landfills or released into the environment through various pathways [8].

Recent studies [2], [9]-[11] suggest that microplastic pollution poses a significant threat to aquatic life, biochemistry of soils, and could possibly be dangerous to human health through the increased microplastic consumption in food and drinking water.

Microplastic detection, quantification and monitoring [12]-[14] in the natural environment, food and air is an essential part of environmental pollution minimization, our understanding of the microplastic pollution mechanisms, key characteristics of micro- and nano-plastic contaminants and their fate in environment.

There are several well-established physical protocols for the analysis of microplastics in the 1 μm to 5 mm particle size range [12]-[15]. These include visual inspection, optical microscopy, Fourier-Transform infrared (FTIR) spectroscopy, Raman/micro-Raman spectroscopy, scanning electron microscopy and X-ray photoelectron spectroscopy. In many situations, especially for smaller microplastic particles with the size less than 200μm, the visual microscopic analysis of microplastics contamination alone is unreliable, leading to inaccurate estimation of microplastic contamination levels [15].

Infrared (IR), optical and X-ray spectroscopic techniques can be applied for accurate microplastics quantification in samples with sub-ppm (parts per million) concentration resolution [16].

However, the application of the optical/IR/X-ray spectroscopic methods is limited to the dedicated laboratory space due to high cost and complexity of equipment, complex procedures of calibration and samples preparation and the necessity in sophisticated software algorithms for spectroscopic data analysis.

In this study, we explore the feasibility of resonance microwave spectroscopy for microplastic detection and quantification in soil and water. Microwave spectroscopic characterization of multi-component materials such as microplastic-contaminated soil or water, is based on the permittivity contrast between the host medium (uncontaminated soil or water) and the microplastic contaminant. This contrast is high, few times difference, in the microplastic-in-soil and extremely high, more than order-of-magnitude difference, in the microplastic-in-water sensing scenarios which creates the opportunity for very accurate microplastic detection and quantification at low concentration levels.

Important features of microwave spectroscopy [17]-[19] include very low cost of microwave sensors, their small size, rugged design for real-time, in-situ operation. These features are attractive for the in-situ microplastic detection and quantification in flexible sea monitoring platforms [20], sewage and wastewater processing plants [21], smart household sensing devices for detecting microplastics in drinking water and food.

The novel linearized model of microplastic concentration quantification in soil and water is developed in this paper and experimentally tested using high-Q resonance printed circuit board (PCB) sensors with simple geometry. The experimental results are obtained for highly-contaminated samples and extrapolated to the lower microplastic concentrations at parts-per-million (ppm) level. The system-level criteria of minimum achievable microplastic concentration resolution in soil and water in terms of the sensor probe Q-factor and microwave receiver sensitivity are formulated in the analytical form and are believed to be applicable to high-Q microwave resonance sensors of arbitrary geometry.

## II. THEORETICAL BACKGROUND

### A. Reflectometry principle of operation and microwave sensor characteristics

A typical microwave reflectometry sensing system, Fig. 1a), consists of a signal source (TX), microwave probe or antenna electromagnetically (EM) interacting with a sample under test, and a receiver (RX) which combines the reflected signal from the probe and a reference signal from the source and generates interferometric output signal. The reflected signal from the probe is characterized by the $S_{11}$ parameter [22] and carries the information about the sample-under-test permittivity. Microwave reflection spectra of the sample are obtained by sweeping the frequency of the source TX and collecting the $S_{11}$ data from which the material parameters (such as microplastic concentration in the sample) can be extracted using suitable mathematical model. The two most important parameters of the microwave reflection spectrum are the resonance frequency defined as the frequency where the magnitude of $S_{11}$ is minimal and Q-factor defined as the ratio of the resonance frequency to the resonance bandwidth at 3dB level [22].

Microwave probes of various geometries can be used for the considered sensing scenario such as microwave cavities or dielectric resonators [23], [24], printed sensors including split-ring [25], [26], microstrip [27], [28] loaded apertures or mini-coil sensors [17]. In this work, a capacitively-loaded PCB square ring, Fig.1b), also called an electric inductive-capacitive resonator [29], is used for the experimental validation.

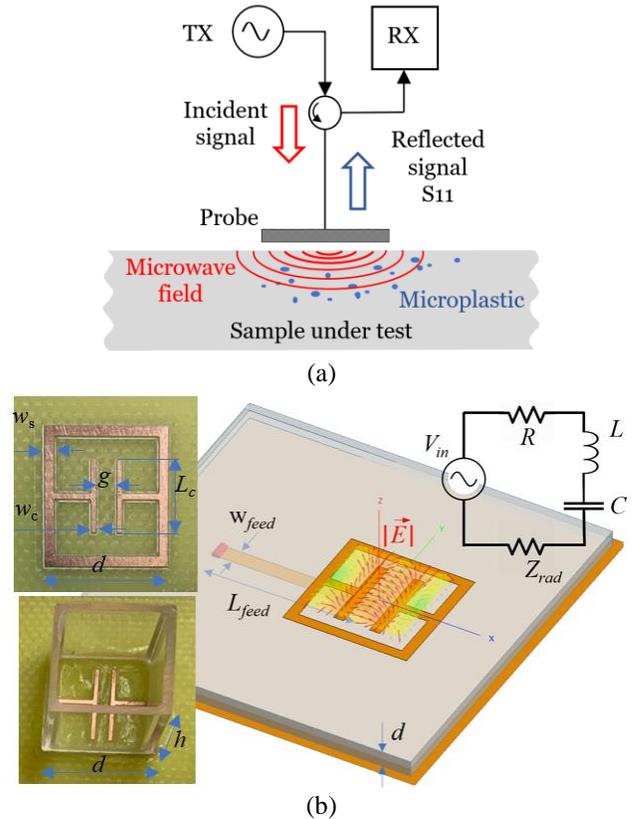

Fig.1. (a) Schematic diagram of the microwave reflectometry setup. (b) printed sensor geometry and integrated soil sample holder; CAD

TABLE I
SIMULATED PERFORMANCE DATA OF THE LOADED LOOP SENSOR

| g (mm) | Resonance frequency (GHz) | Q-factor | $\|E\|$ z =1mm (V/m) | $\|E\|$ z =10mm (mV/m) | Rad. gain (dBi) |
|---|---|---|---|---|---|
| 0.5 | 5.40 | 223.1 | 0.91 | 88.64 | 3.03 |
| 1.0 | 5.57 | 124.6 | 1.00 | 97.35 | 3.08 |
| 1.5 | 5.71 | 95.2 | 0.98 | 102.87 | 3.12 |
| 2.0 | 5.83 | 87.1 | 0.91 | 105.18 | 3.14 |
| 2.5 | 5.98 | 89.6 | 0.80 | 104.74 | 3.16 |
| 3.0 | 6.10 | 93.4 | 0.67 | 99.22 | 3.18 |

model, simplified equivalent circuit model (top corner inset) and simulated *E*-field distribution in the proximity (at 1mm stand-off distance) to the sensor aperture.

The reason for this choice is i) simple, fully planar sensor geometry, ii) *E*-field collimation in the vicinity of the sensor aperture inside the capacitive gap and around the capacitive plates, Fig1b). This field collimation is essential for efficient EM coupling between the sensor and the sample-under-test material. iii) high Q resonance, Q-factor is more than 200, iv) the possibility of the resonance frequency control by a small change in the gap dimensions which can be employed for sensor-array based smart sensing [29], and v) moderate far-field radiation gain, Table I, which can be used for the remote RFID interrogation of the immersed-into-soil sensor response in the in-situ sensing scenario.

Table I summarizes FEKO-simulated EM parameters of the sensor for the dimensions $d$ =12mm, $L_c$ =8mm, $w_s$ = $w_c$ = 1mm, $L_{feed}$ =40mm, $w_{feed}$=1.5mm and gap $g$ being changed from 0.5mm to 3mm with 0.5mm step. The FR4 substrate thickness is 1.6mm, the feeding strip is vertically separated at 0.8mm from the ground plane and the sensor aperture. These simulated parameters are in good agreement with the experimental data, Section III.

### B. Mathematical model of the $S_{11}$ resonance frequency shift

The resonance frequency shift $\Delta f = |f_0 - f_c|$ between the $S_{11}$ spectral lines of the uncontaminated sample (resonance frequency $f_0$) characterized by the relative permittivity $\varepsilon_r^{(h)}$ and the spectral line of the contaminated sample (resonance frequency $f_c$) with relative permittivity $\varepsilon_r^{(mix)}$,

$$\varepsilon_r^{(mix)} = V_h \varepsilon_r^{(h)} + V_c \varepsilon_r^{(c)} \quad (1)$$

can be written in the analytical form as (Appendix A)

$$\Delta f = \frac{1}{2} f_0 V_c \frac{\varepsilon'^{(h)}_r - \varepsilon'^{(c)}_r}{A + \varepsilon'^{(h)}_r} \quad (2)$$

In (1), $V_h$ and $V_c$ are the volumetric concentrations (per unit volume) of the host and contaminant material, respectively $V_h + V_c = 1$. Formula (1) describes the relative permittivity of two-component mixture composed of the host medium and contaminant when $V_c \ll V_h$, [30], [31].

In (2), $\varepsilon'^{(h)}_r$ and $\varepsilon'^{(c)}_r$ are the real parts of the complex-valued relative permittivity of the host medium and contaminant respectively. $A$ is a constant that depends on the sensor geometry [32] and, in general permittivity of the sample, and can be found from comparing the reflection spectra of an empty (unloaded) sensor and a sensor loaded with a *calibrated* sample under test, (e.g. clean soil). Detailed derivation of equation (2) is provided in Appendix A.

It is important to note that for small contaminant concentrations, $V_c < 0.1 V_h$, the resonance frequency shift $\Delta f$ is linearly proportional to the contaminant concentration $V_c$ and the permittivity contrast $\varepsilon'^{(h)}_r - \varepsilon'^{(c)}_r$. Since the permittivities of the host medium (e.g. water or soil) and microplastic contaminant can be easily measured or obtained from the reference data, equation (2) forms the analytical basis to quantify microplastic concentrations, by measuring the resonance frequency shift of the microwave signal, reflected from the sample under test.

### C. Theoretical limits of microplastic concentration resolution

From (2), it can be seen that the microplastic volumetric concentration in the host medium

$$V_c = 2 \frac{\Delta f}{f_0} \frac{A + \varepsilon'^{(h)}_r}{\varepsilon'^{(h)}_r - \varepsilon'^{(c)}_r} \quad (3)$$

is linearly proportional to the relative frequency shift $\Delta f/f_0$, for the effective permittivity linear approximation (1). Quantification of the microplastic concentration in the host material from the $S_{11}$ spectra is based on detection of the resonance frequency of an uncontaminated sample (calibration sample) and relative resonance frequency shift of the contaminated sample spectral line. Resonance frequency measurement is based on the $S_{11}$ minimum detection, which poses certain requirements of the electronic circuitry characteristics. To understand these requirements better, one can derive an expression for the $S_{11}$ parameter magnitude as a function of frequency $f$ in the vicinity of the resonance frequency point $f_0$,

$$|S_{11}(\Delta f)| = |S_{11\,min}| \sqrt{1 + 4Q^2 \frac{\Delta f^2}{f_0^2}} \quad (4)$$

Where $\Delta f = |f - f_0|$, $\Delta f/f_0 \ll 1$. $|S_{11\,min}|$ is a magnitude of $S_{11}$ parameter at the resonance $f_0$. Q is a quality factor of the sensor probe. The derivation of (4) is detailed in Appendix B. It should be stressed that (4) is derived under general reflection resonance conditions (Appendix B) and is not specific to the particular sensor type used in this work.

Using Taylor series expansion in (4) for the small parameter $4Q^2 \Delta f^2/f_0^2 < 0.1$, it is possible to find the $S_{11}$ magnitude variation $|\Delta S_{11}|$ between the contaminated and uncontaminated material spectral lines at the frequency $f_0 + \Delta f$

$$|\Delta S_{11}| = |S_{11}^{(c)}(f_0 + \Delta f) - S_{11}^{(0)}(f_0 + \Delta f)| \quad (5)$$

in the analytical form

$$|\Delta S_{11}| = 2|S_{11\,min}| \left(Q \frac{\Delta f}{f_0}\right)^2 \quad (6)$$

In (5) $f = f_0 + \Delta f$ is the resonance frequency of the contaminated sample reflection spectral line, $S_{11}^{(c)}(f)$ and $S_{11}^{(0)}(f)$ are the spectral lines of contaminated and uncontaminated samples and it is assumed that $|S_{11\,min}|$ does not appreciably change with a small frequency offset $\Delta f$.

On the other hand, from (3) it is possible to write the contaminant volumetric concentration resolution as

$$\Delta V_c = D \, \Delta f / f_0 \quad (7)$$

where constant $D = 2(A + \varepsilon'^{(h)}_r) / (\varepsilon'^{(h)}_r - \varepsilon'^{(c)}_r)$.

Equations (6), (7) specify the best achievable resolution, in the linear detection regime, of the microplastic contaminant in the soil or aquatic environment:

i) to achieve the required $\Delta V_c$ resolution, the frequency tuning resolution (frequency step) $\Delta f/f_0$, of the source TX and receiver RX

ii) should be finer than $\Delta V_c/D$;

ii) ii) for the specified frequency step $\Delta f/f_0$, the microwave receiver RX should possess the amplitude sensitivity better than (6).

The required signal-to-noise ratio $S/N$ at the receiver output is related to the sensitivity (6) by an equation [33]-[35]

$$|\Delta S_{11}| = kTB \, S/N \qquad (8)$$

Where $k$ is the Boltzmann's constant, $T$ is the equivalent noise temperature of the receiver referenced to its input and $B$ is the RX bandwidth, $|\Delta S_{11}|$ dimension can be converted to W assuming 50Ω terminal impedance.

The typical frequency tuning resolution of the phase-locked-loop microwave transmitters and receivers [36], [37] is in the range $10^{-5} – 10^{-8}$, which indicates that the theoretical microwave spectroscopic resolution of microplastic concentrations in the environment at tens of ppm level is possible, considering microwave circuitry frequency tuning characteristics.

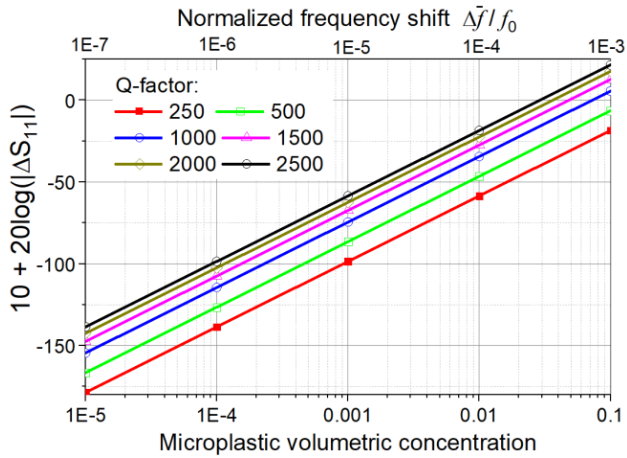

Fig.2. Required receiver sensitivity defined as $10 + 20 \log(|\Delta S_{11}|)$ as a function of microplastic-in-soil volumetric concentration (per unit volume).

Fig.2. shows the required RX sensitivity for the microplastic-in-sand detection using estimation model (2)-(6). The equation parameters in (2), $\varepsilon'^{(h)}_r = 5.60, \varepsilon'^{(c)}_r = 2.59, A = 55.8$ are obtained from the experimental results in Section III. The Q-factor varies from 250 (sensor in this work) to 2500. From Fig.2 it can be seen that higher Q factors of the sensing probe allow better contaminant resolution, however even for Q=2500, the concentration resolution is limited to 100ppm (at the RX sensitivity -100dBm). Several strategies can be employed to increase the contaminant concentration resolution, including low-noise signal amplification [18], [19] and/or sensing volume reduction [38], thus effectively increasing contaminant volumetric concentration.

## III. Experimental Results

In this section, experimental data demonstrating microplastic detection and quantification in soil and in fresh and salty water are presented. The soil and water samples were artificially contaminated with large concentrations of microplastic, at tens to hundreds of parts-per-thousand (ppt) level, to achieve accurate experimental control of the contaminant concentration levels in the host material.

### A. Experimental Detection and Quantification of Microplastic in Soil

*Soil samples*

Two types of clean soil were prepared: topsoil with high (approximately 20%) percent of organic matter (pH is around 7%) and bulk density of 1.33g/cm³ and sandy soil with high (more than 90%) of medium-coarse sand content, bulk density 2.56g/cm³. The topsoil and sandy soil samples were collected from field around the GPS locations (54.556745, -5.928940) and (54.337611, -5.837946), respectively, in Northern Ireland and dried in the oven at 90°C for 20 minutes. The topsoil samples were further manually sieved thorough a 0.5mm mesh to remove large organic debris and larger particles to homogenize the soil. Both samples are shown in Fig.3a),b).

*Plastic samples*

The microplastic is Nylon 11 (Rilsan ®) powder, with density of 1.03g/cm3, Fig.3c). The microplastic powder particles size range is 50μm-0.5mm. The complete mechanical and dielectric properties are available elsewhere [39] for the bulk Nylon 11 material.

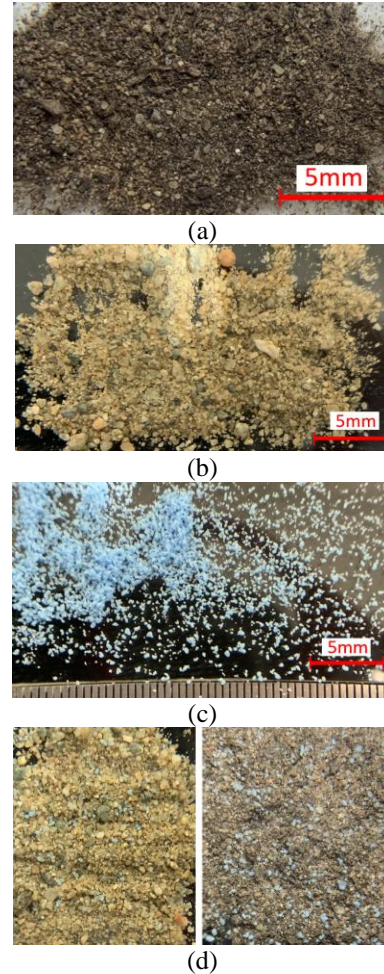

Fig.3. Samples photographs: (a) topsoil, (b) sandy soil, (c) Nylon 11 powder, (d) microplastic-contaminated samples: left- 10ppt microplastic-in-sand, right – 50ppt, microplastic-in-soil.

This microplastic is chosen due to its global production volume, and wide applications area, especially in 3D printing [40].

The samples complex permittivity was measured in 5.0-6.0 GHz band using conventional free-space measurement method [24]. The measured permittivity values are summarised in Table II for the frequency 5.5GHz. The experimental $|S_{11}|$ spectra of the microplastic- contaminated samples are shown in Fig. 4.

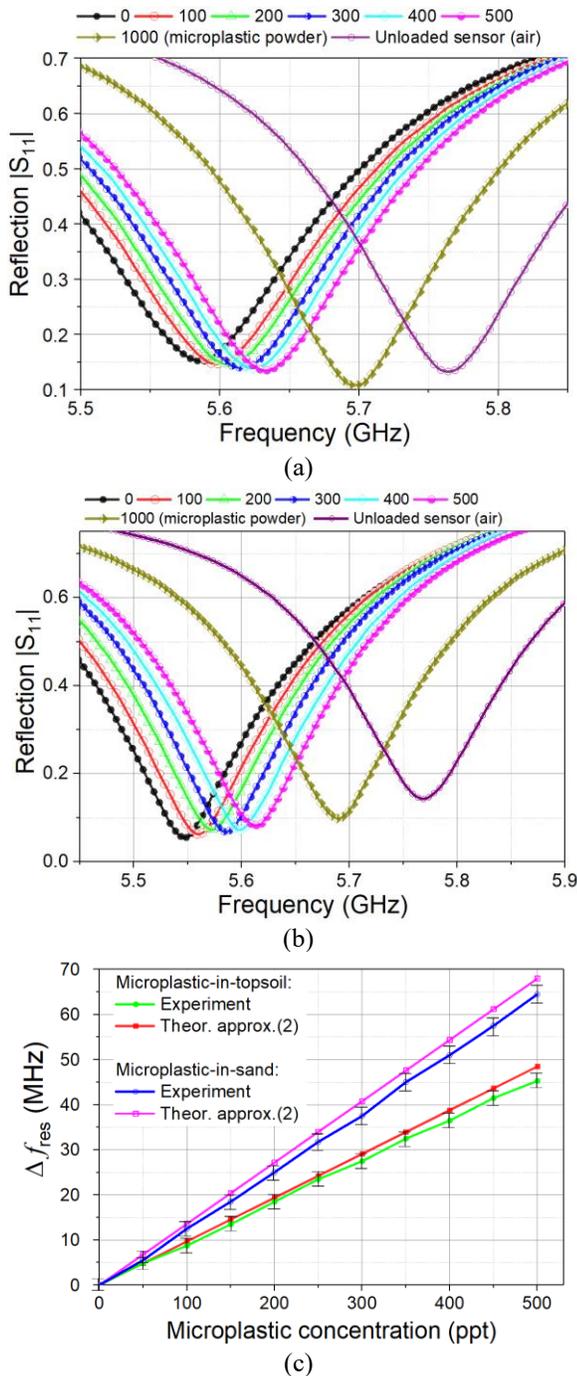

Fig.4. Measured $S_{11}$ spectra of the microplastic-contaminated topsoil (a); and contaminated sandy soil (b). $S_{11}$ resonance frequency shift vs microplastic concentration (c). Sensor aperture dimensions are $d$ =12mm, Lc =8mm, ws = wc = 1mm, Lfeed =40mm, wfeed=1.5mm, g=1.5mm. The sensor is integrated with acrylic sample holder of 12mmx12mm x10mm dimensions and wall thickness 1mm.

The Keysight E8361C performance network analyzer was used to measure the $S_{11}$ data. It was found that the resonance frequency shift is sensitive (up to ±3MHz) to the sample particles distribution in the contact with the sensor aperture, which can be explained by a non-uniform spatial distribution of the $E$-field, Fig.1b).

To achieve accurate samples characterization, ten sub-samples of each microplastic concentration from 0 to 500ppt with 50ppt step were prepared. These sub-samples were thoroughly mixed and $S_{11}$ measured for each sub-sample resulting in the resonance frequency sub-bands ($f_{min,i}$, $f_{max,i}$), index $i$ =1,2,..,10, corresponds to 50ppt concentration step. Each $S_{11}$ spectral line shown in Fig. 4 a), b) corresponds to the center of the resonance frequency sub-band. Effectively, this measurement procedure is equivalent to spatial averaging of the contaminated samples particles distribution and the sensor $E$-field non-uniform spatial features.

TABLE II
MEASURED PERMITTIVITY OF DRY SOIL AND NYLON 11 POWDER AT 5.5GHZ

| Material | Permittivity real part | Loss tangent |
|---|---|---|
| Topsoil | 4.70 ± 0.20 | 0.03 ± 0.002 |
| Sandy soil | 5.60 ± 0.20 | 0.007 ± 0.002 |
| Nylon 11 powder | 2.59 ± 0.20 | 0.008 ± 0.002 |

The measured Q-factors of the empty and sample-loaded sensors are summarized in Table III. These Q-factor values can be explained using the permittivity data, Table II which shows that sandy soil has the largest real-part permittivity and the lowest microwave loss factor. The lower Q-factor of the topsoil-loaded sensor is caused by the higher tan δ of the topsoil sample.

TABLE III
Q-FACTORS OF EMPTY AND LOADED SENSORS

| Sample | Q-factor |
|---|---|
| Air | 119.4 |
| Uncontaminated sandy soil | 259.8 |
| Uncontaminated topsoil | 92.4 |
| Nylon 11 powder | 157.4 |

Finally, constant $A$ appearing in (2), (3) was obtained from the measurement data, Fig. 4a), b) by comparing the unloaded and loaded sensor resonance frequency, $A$ =55.8.

Fig. 4c) shows the measured resonance frequency shift vs microplastic concentration in topsoil and sand samples and the resonance frequency shift given by a linearized equation (2). The error bars show the resonance frequency sub-bands ($f_{min,i}$, $f_{max,i}$). It can be seen that the linear approximation (2) is in a very good agreement with the experimental data, especially at lower concentration of the microplastic contaminant in soil.

### B. Experimental Detection and Quantification of Microplastic in Water

In the microwave frequency band 1GHz-10GHz, the permittivity and microwave loss contrast between water and most plastics differ by more than order of magnitude. Specifically, at 5.5GHz the permittivity and loss tangent of fresh and sea (30ppt) water obtained from the literature [41]-[43] are listed in Table IV suggesting that there is more than 20 times contrast in the real part of permittivity and two orders of magnitude contrast in loss tangent between water and Nylon 11 microplastic powder.

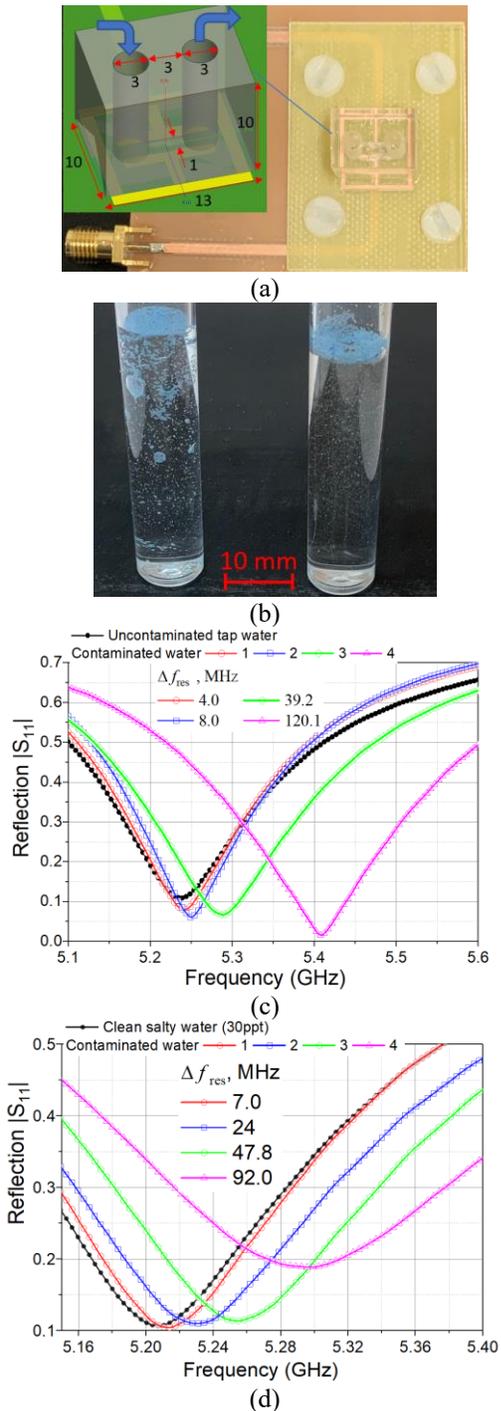

Fig.5. Square loop capacitive resonator, partially-loaded with water-filled channel, inset shows cavity dimensions in mm, (a). Photograph of the microplastic-water mixture, left- tap water, right- salty water, (b). Measured S11 spectra of the microplastic-contaminated tap, (c) and salty, (d), water.

Sensing in aquatic environment has a number of distinct features, dictated by the high permittivity and microwave loss of water [44]-[46]. In this paper, it was experimentally found that full immersion of the microwave sensor into tap or salty water leads to considerable reduction of the Q-factor (5 times or more). To maintain high Q essential for accurate contaminant detection, the inductive-capacitive square-ring sensor is *partially* loaded with a water mini-channel of 1mm diameter and 9mm length, Fig. 5a), drilled in a miniature acrylic block. This water-filled mini-channel is positioned between the capacitive plates of the sensor, as shown in Fig.5a).

TABLE IV
COMPLEX RELATIVE PERMITTIVITY OF WATER @5.5GHZ AND 20°C

| Sample | Re $\varepsilon_r$ | Loss tangent |
|---|---|---|
| Fresh water | 72.0±0.50 | 0.28±0.02 |
| Salty water | 66.0±0.40 | 0.80±0.07 |

Partial loading of the resonator allows to preserve high Q factor: $Q_{air}$ =218.7 (sensor with empty acrylic cavity), $Q_{tap\_water}$ = 110.0 (uncontaminated tap water), $Q_{salty\_water}$ =101.5 (uncontaminated salty water). These data show that Q factor is reduced by two times as compared to the Q-factor of the sensor loaded with low-loss sand samples.

Another important feature of microplastic-in-water detection is related to fluid dynamics of microplastic-in-water mixture, Fig.5b). Nylon 11 is hydrophobic material (water absorption index 0.8% over 24 hours), so it is resistant to solubility with water. In the samples shown in Fig.5b), 50ppt volume of Nylon 11 powder is mixed with tap (left) and salty (right) water. In tap water, microplastic forms emulsion-like mixture of suspended particles (less than 100μm) and mini-clusters of particles of 0.25mm-2mm average size. In salty water, the Nylon particle clusters are formed only on the water surface, the distribution of particles inside water volume is more uniform. Since microplastic particles size distribution can vary from approximately 50μm to 0.5mm, accurate quantification of the $S_{11}$ spectral lines vs particle size distribution requires microsystem design of mesh filters, mini- or microfluidic channels and a resonator array, measuring $S_{11}$ for specific microplastic particle size distribution, which will be reported in future work.

In this study, preliminary experimental results demonstrating characteristic resonance frequency variation in microplastic-contaminated samples are presented in Fig. 5c),d). The overall range of resonance frequency shift was more than 140MHz, depending on the Nylon 11 particles distribution in water mini-channel. The water flow was controlled by a syringe pump, particle size distribution was monitored under the microscope and $S_{11}$ measured for specific particle distribution in the mini-channel. In Fig. 5c) the approximate linear size of Nylon 11 mini-clusters in the channel: line 1 - 0.2mm, 2 – 0.3mm, 3 – two clusters x 0.3mm, line 4 – clusters 0.5mm and 0.6mm. Fig.5d) line 1- 0.25mm, line 2 – two clusters x 0.3mm, line 3- two clusters 0.35mm and 0.45mm, line 4 – two clusters 0.6mm and 1mm. These data show that the resonance frequency shift increases proportionally to the volumetric concentration of microplastic in water, which is in a good qualitative agreement with the model (2).

To summarize, it is demonstrated that resonance microwave sensing is a promising tool to detect and quantify microplastic in soil and water at ppt concentration levels. The theoretical model (2) developed in this study, suggests that the resonance frequency shift of the $S_{11}$ spectral line of a contaminated sample is linearly proportional to the volumetric concentration of microplastic in the host medium. Estimation analysis (3)-(8) demonstrates the limitations of microwave sensing at the contaminant concentration levels below 100ppm, due to required sensitivity of the receiver. These limitations can be overcome by increasing the Q-factor of the sensor, low-noise signal amplification, non-linear signal detection and reducing the sensing volume of the sample under test.

## APPENDIX A

The resonance frequency $f_L$ of a microwave sensor loaded with a material sample with real-part relative permittivity $\varepsilon'_r$, and the resonance frequency $f_U$ of an unloaded sensor (air) are connected by the relation [32]

$$f_L^2/f_U^2 = (1 + A)/(\varepsilon'_r + A) \quad (A1)$$

where $A$ is a constant that depends on the sensor geometry and sample permittivity $\varepsilon'_r$. If the host medium (soil or water, $\varepsilon'^{(h)}_r$) is contaminated with microplastic ($\varepsilon'^{(c)}_r$) such that the permittivity of the mixture

$$\varepsilon'_r = \varepsilon'^{(h)}_r - \Delta\varepsilon \quad (A2)$$

where $\Delta\varepsilon << \varepsilon'^{(h)}_r$ is a small permittivity variation,

$$\Delta\varepsilon = \varepsilon'^{(h)}_r - \left(V_h \varepsilon'^{(h)}_r + V_c \varepsilon^{(c)}_r\right) = V_c\left(\varepsilon'^{(h)}_r - \varepsilon'^{(c)}_r\right) \quad (A3)$$

one can write the expression for the resonance frequency of the sensor loaded with a contaminated sample as

$$f_L = \frac{f_U\sqrt{1+A}}{\sqrt{A+\varepsilon'^{(h)}_r - \Delta\varepsilon}} = \frac{f_U\sqrt{1+A}}{\sqrt{A+\varepsilon'^{(h)}_r}\sqrt{1-\frac{\Delta\varepsilon}{A+\varepsilon'^{(h)}_r}}} \quad (A4)$$

Taking into account that $\Delta\varepsilon/(A + \varepsilon'^{(h)}_r) < 0.1$ and using Taylor expansion for inverse square root in (A4), one can derive the equation for the resonance frequency $f_L$ of the sample with microplastic contaminant volumetric fraction $V_c$,

$$f_L = f_0 + \frac{1}{2} f_0 \frac{V_c\left(\varepsilon'^{(h)}_r - \varepsilon'^{(c)}_r\right)}{A+\varepsilon'^{(h)}_r} \quad (A5)$$

where $f_0$ is the resonance frequency of the sensor loaded with uncontaminated sample, $V_c = 0$. Equation (2) follows from (A5).

## APPENDIX B

The magnitude of the $S_{11}$ parameter of a resonance sensor can be written in a general form [22] as

$$|S_{11}| = |(Z_L - Z_0)/(Z_L + Z_0)| \quad (B1)$$

where $Z_L = R + j(\omega^2 LC - 1)/\omega C$ is a complex impedance of a resonance sensor, $R$, $L$, $C$ are the equivalent lumped circuit parameters, $Z_0$ is an impedance of the sample material. In the proximity to the resonance, $f = f_0 \pm \Delta f$, the real part of the denominator dominates over the imaginary part and almost constant, the numerator can be represented in the form

$$Z_L - Z_0 = Re\,\Delta Z + jL(\omega^2 - \omega_0^2)/\omega \quad (B2)$$

Expanding $\omega^2 = 4\pi^2(f_0 + \Delta f)^2$, neglecting higher order terms $\sim \Delta f^2$ and taking the module of (B2), one can derive the approximation for the $|S_{11}|$ magnitude around the resonance frequency $f = f_0 + \Delta f$,

$$|S_{11}(f)| \approx |S_{11min}||(1 + 4Q^2\Delta f^2/f_0^2)^{1/2}| \quad (B3)$$

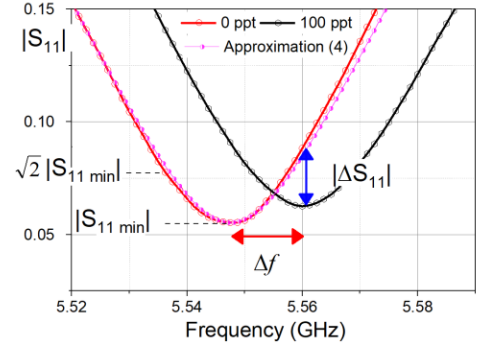

Fig A1. Reflection parameter $|S_{11}|$: red dotted line – measurement, magenta solid line- approximation. The approximation (B3) is in excellent agreement with the measured $|S_{11}|$ data.

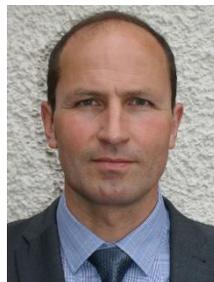

Oleksandr Malyuskin is a Senior Lecturer (Associate Professor) at the Centre for Wireless Innovation, ECIT, Queen's University Belfast. His research interests are focused on microwave imaging and sensing for industrial, environmental and biological applications, advanced wireless communications including time-reversal, retrodirective and electrically-small antenna arrays, electromagnetic metasurfaces and metamaterials. He is currently involved, as a leading researcher in a number of projects aimed at the nano-materials and biomaterials characterization and nanomaterials-based sensing. He is a co-author of more than 50 IEEE publications in the field of microwave imaging, sensing and wireless communications.